# Quantum Machine Learning for Health State Diagnosis and Prognostics

Gabriel San Martín Silva

Department of Civil and Environmental Engineering, University of California Los Angeles, USA. E-mail: gsanmartin@g.ucla.edu

Enrique López Droguett, Ph.D.

Department of Civil and Environmental Engineering, and Garrick Institute for the Risk Sciences, University of California Los Angeles, USA. E-mail: eald@ucla.edu

*ABSTRACT*

Quantum computing is a new field that has recently attracted researchers from a broad range of fields due to its representation power, flexibility and promising results in both speed and scalability. Since 2020, laboratories around the globe have started to experiment with models that lie in the juxtaposition between machine learning and quantum computing. The availability of quantum processing units (QPUs) to the general scientific community through open API's (e.g., IBM's Qiskit) have kindled the interest in developing and testing new approaches to old problems. In this paper, we present a hybrid quantum machine learning framework for health state diagnostics and prognostics. The framework is exemplified using a problem involving ball bearings dataset. To the best of our knowledge, this is the first attempt to harvest and leverage quantum computing to develop and apply a hybrid quantum-classical machine learning approach to a prognostics and health management (PHM) problem. We hope that this paper initiates the exploration and application of quantum machine learning algorithms in areas of risk and reliability.

## 1 INTRODUCTION

Deep learning has been used for almost a decade now by the prognosis and health management community to address questions such as the identification of health states of rotation machinery [1], the prediction of remaining useful life [2] or Bayesian analysis within complex and connected systems [3]. Since 2015, commercial applications have also appeared, with a clear intention of leveraging the hidden value contained within the immense amount of operational data that companies have collected through the years. Examples of this can be found through the specialized literature [4], [5]. While both research and commercial applications have produced useful results within the context of big data and complex, multidimensional systems (e.g., the assessment of health states through the analysis of images or videos) there are challenges that need to be addressed before these applications can be safely implemented in safety critical operations. One of them is speed and scalability: even with deep learning, for extremely dense or complex models, their training and inference can be quite challenging to execute, even more so to deploy on site. Over the last two years, general media and the research community have been starting to pay attention to advances in the quantum computing field, motivated mainly by remarkable feats such as the quantum supremacy experiment [6] and the first hardware ready Noisy Intermediate-Scale Quantum (NISQ) computers becoming available to the general public through cloud services, with hopes of identifying possible ways to optimize and speed up existing algorithms, modifying them or developing novel ones that harvest quantum mechanics phenomena such as entanglement, superposition and interference [7]. In this context, we present what it is, to the best of our knowledge, the first attempt to develop and apply a hybrid quantum-classical machine learning algorithm to address PHM problems, more specifically, diagnosis of health states in rotary machinery. We also present and discuss the encoding schema used to convert classical health monitoring data into quantum data and injected into quantum circuits, a required step to process information within the quantum computing realm. We will also release the code in a public GitHub repository required to reproduce the experiments and results portrayed here.

The rest of the paper is organized as follows. Section 2 presents a brief introduction to fundamental quantum computing concepts, such as qubits, quantum gates, encoding schemas and a high-level view on quantum machine learning. Section 3 follows with the description of the hybrid approach for tackling health diagnosis and prognosis problems. Section 4 describes the case study: both the dataset used, the model specification and the results obtained. We conclude with a brief overview of the main takes of this work in Section 5.

## 2 THEORETICAL BACKGROUND

### 2.1 Qubits and Quantum Gates

Within the world of classical computation, a bit is the basic unit of information, only capable of representing the state of a system which can be in one of two states. Almost every computational system that we used in our everyday lives can be decomposed into bits and logical operations between bits (also called gates), with varying levels of abstraction. Similarly, quantum computing also has these basic components, called quantum bits (qubits) and quantum gates.

As with traditional bits, qubits also represent the state of a system with two states. Nevertheless, the quantum system is not restricted to be in one of these two states. This means that the system can be in both states at the same time or, more precisely, a complex linear combination of the two basis states, a fundamental quantum mechanics property known as superposition. Mathematically, a qubit is described using a 2-by-1 matrix with complex numbers as its components:

$$|\psi\rangle = [c_0, c_1]^T \tag{1}$$

where $|c_0|^2 + |c_1|^2 = 1$, i.e., probability axiom holds. When we measure a qubit, it automatically collapses to a bit, with probability $|c_0|^2$ that it collapses to the state $|0\rangle$ and $|c_1|^2$ to the state $|1\rangle$ hence why the sum of the squared modulus of both complex numbers must add-up to 1. Useful systems are composed of multiple qubits, obtained through the tensor product to represent all the possible combinations of states:

$$|\psi\rangle = |\psi_1\rangle \otimes |\psi_1\rangle \otimes \ldots \otimes |\psi_N\rangle \tag{2}$$

It is easy to note that a system composed of $N$ qubits can represent, simultaneously, $2^N$ different states, while with classical bits there exists $2^N$ possible states, but the system is always fixed in one of them. This feature is what fundamentally allows quantum computing to perform certain tasks with a significant theoretical advantage [8].

Quantum Gates represent the basic operations that can be performed on a qubit or a group of qubits. Mathematically, they are described by unitary matrices to assure the reversibility of the operation. In what follows, we describe the most common gates in quantum computing.

*Hadamard Gate (H)*

This gate is used to induce a qubit into a superposition state. Applied over $|\Psi_0\rangle = |0\rangle$, its output corresponds to a qubit that has an equal probability of collapsing after measurement to a $|0\rangle$ or $|1\rangle$ state:

$$H = \frac{1}{\sqrt{2}} \begin{bmatrix} 1 & 1 \\ 1 & -1 \end{bmatrix} \tag{3}$$

*Controlled-Not Gate (C-NOT)*

This gate is used to entangle (make the state of a qubit dependent on the state of the other) two qubits. This gate has two inputs and two outputs. One input qubit act as the control qubit, while the other qubit acts as the controlled one. If the control qubit after the measurement process is $|0\rangle$, then the controlled qubit remained untouched. On the other hand, if the control qubit after the measurement process collapses to a $|1\rangle$ state, then the controlled qubit is inversed by this gate:

$$C - NOT = \begin{bmatrix} 1 & 0 & 0 & 0 \\ 0 & 1 & 0 & 0 \\ 0 & 0 & 0 & 1 \\ 0 & 0 & 1 & 0 \end{bmatrix} \tag{4}$$

*Bloch Sphere Representation and Rotation Gates*

As we showed before, a single qubit is composed of a linear combination of two complex numbers ($c_0$ and $c_1$) and two basis vectors ($|0\rangle$ and $|1\rangle$). Using the polar representation of complex numbers, we can represent a qubit as:

$$|\psi\rangle = r_0 e^{i\phi_0}|0\rangle + r_1 e^{i\phi_1}|1\rangle \tag{5}$$

While this representation shows that a qubit is composed of 4 real variables ($r_0, r_1, \phi_0, \phi_1$) using the facts that $|c_0|^2 + |c_1|^2 = 1$ and that the quantum state does not change if multiplied by a number of unit norm (for example, $e^{-i\phi_0}$) [9], it can be shown, through the following operations, that the qubit only has 2 free parameters :

$$|\psi\rangle = e^{-i\phi_0}(r_0 e^{i\phi_0}|0\rangle + r_1 e^{i\phi_1}|1\rangle) \tag{6}$$

$$|\psi\rangle = r_0|0\rangle + r_1 e^{i(\phi_1 - \phi_0)}|1\rangle \tag{7}$$

$$|\psi\rangle = \cos\theta\, |0\rangle + \sin\theta\, e^{i\varphi}|1\rangle \tag{8}$$

Where we have used the unit norm of the original qubit to replace $r_0$ and $r_1$ for the single free parameter $\theta$. Given that the qubit can be represented by two angles, $\theta$ and $\varphi$, we can also provide a graphical representation of it through a vector encapsulated within a sphere or unit radius. This sphere is often known as the Bloch sphere:

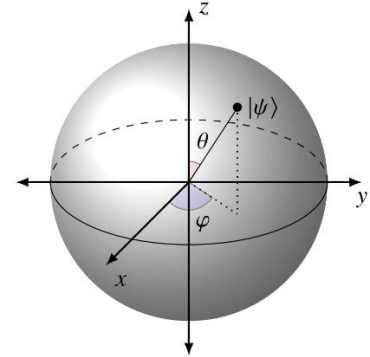

*Figure 1: The Bloch Sphere as a graphical representation of a single qubit. Note when $\theta = 0$, the representation of $|\psi\rangle$ aligns with the z-axis, which represents the $|0\rangle$ state. The case when $\theta = \pi/2$ is particularly interesting, since it represents a case in which both basis states, $|0\rangle$ and $|1\rangle$, form a complex linear combination, and therefore the qubit is in a perfectly balanced superposition state, with equal probability to collapse to any of those two states. Figure from [7].*

The final three gates that we will use in the topology of the proposed approach are the rotational gates, which generates a rotation of the qubit of $\xi$ radians around the X, Y and Z axis of the Bloch sphere, i.e.:

$$R_x(\xi) = \begin{bmatrix} \cos\frac{\xi}{2} & -i\sin\frac{\xi}{2} \\ -i\sin\frac{\xi}{2} & \cos\frac{\xi}{2} \end{bmatrix} \tag{9}$$

$$R_y(\xi) = \begin{bmatrix} \cos\frac{\xi}{2} & -\sin\frac{\xi}{2} \\ \sin\frac{\xi}{2} & \cos\frac{\xi}{2} \end{bmatrix} \quad (10)$$

$$R_z(\xi) = \begin{bmatrix} e^{-i\frac{\xi}{2}} & 0 \\ 0 & e^{i\frac{\xi}{2}} \end{bmatrix} \quad (11)$$

### 2.2 Encoding Schemas

One of the main challenges when working with quantum circuits and algorithms is to encode real-valued data into qubits. This task, apparently trivial, is not easy and is nowadays a highly active area of research. Currently, the simpler encoder schema is called Angled Encoding [10], and it works generating a circuit that takes as input $N$ qubits set in a $|0\rangle$ state and a $N$-dimensional real valued vector, to output the system of $N$ qubits, but with the classical data encoded into their phases.

For this, the operation that is performed is as follows:

$$\vec{x} \to |\psi\rangle = S(x_0) \otimes S(x_1) \otimes \ldots \otimes S(x_{N-1}) \quad (12)$$

where $S$ represents the following operation, performed to each element of the classical vector:

$$S(x_i) = \cos\left(\frac{\pi}{2}x_i\right)|0\rangle + \sin\left(\frac{\pi}{2}x_i\right)|1\rangle \quad (13)$$

Therefore, this encoding schema uses $N$ qubits to encode a $N$-dimensional vector which values lie in the $[0,1]$ interval. While other encoding schemas are more efficient in terms of memory, Angle Encoding is simple and performant, so it is used in the case study discussed in Section 4.

### 2.3 Quantum Machine Learning

Quantum Machine Learning is a new field that attempts to use quantum computing techniques to develop new algorithms and improve or speed up traditional machine learning algorithms. As with any field, research has branched out looking for multiple ways to achieve this objective. During the last five years, researchers have attempted to inject quantum computing algorithms into specific parts of classical machine learning algorithms, for example to accelerate linear algebra computations [11] or to propose new ways of performing optimization [12]. This is commonly known as first phase quantum machine learning [13]. While, in theory, these fused algorithms have shown performance advantages, the scalability of these techniques is still under debate since modern-day Quantum Processing Units (QPUs) cannot reliably handle the computations required to test these approaches on real data. An alternative approach to this is the use of small, parameterized quantum circuits, which can be easily executed on real hardware or even simulated in classical computers to study the impact of adding quantum circuits as building blocks to new frameworks. In a sense, this way of approaching this new field is similar to what took place around 2010 with deep learning models as a natural evolutional of machine learning models to achieve more scalability. Again, the quantum machine learning community often refers to this approach as the second phase.

In this paper, we make use of a second phase approach, developed initially by the TensorFlow Quantum team and proposed alongside its library for quantum computing [13].

### 2.4 Potential Applications to RAMS

In general, quantum computing can be directly applied to RAMS tasks through the enhancement of existing machine learning models, for the diagnosis of health states or the prognostics of important magnitudes such as a damage level or remaining useful life, where the use of entangled and superposed qubits could yield important improvements in both computing speed and representation efficiency.

Nevertheless, quantum computing could also be applied in a more indirect manner. One example is the use of new sampling techniques, such as [14], to improve probabilistic risk and reliability assessment, in which the limiting factor is obtaining enough samples to successfully represent, or model a given system.

A second example of a possible application is the use of quantum optimization techniques for any RAMS related tasks that need a faster, more computational efficient way of performing optimization. For example, the quantum approximation optimization algorithm (QAOA) [15] is designed to solve combinatorial optimization problems and could be used for spare parts inventory management.

## 3 PROPOSED APPROACH

The proposed approach is composed of four basic building blocks. First, there exists a traditional preprocessing data step. In the general case, this could be the normalization of images to a color-map range, or the reduction of dimensionality using principal component analysis, for example, or raw time-series monitoring data such as multi-process sensors (e.g., flow, temperature, pressure) and vibration monitoring as well as frequency and time-dependent features as a way of reducing dimensionality. Then, the processed classical data is converted to quantum circuits containing the original information using an encoding schema, explained generally in Section 2.2. Next, the quantum encoded data is fed to a Parameterized Quantum Circuit (PQC), which is a user defined quantum circuit composed of multiple gates that can accept external, free parameters, and that will be later optimized in the training phase. After the PQC, we introduce a measurement step, which effectively returns the now processed quantum data to classical data, which is fed to the last step of the hybrid approach, namely a classical (potentially deep) neural network that will perform the required fault diagnosis and prognosis. Figure 2 shows a schematic view of the hybrid quantum-classical approach. As portrayed in Figure 2, the backpropagation technique is used to evaluate gradients and update parameters in both the PQC and classical neural networks. For the PQC, this can be achieved using various techniques, such as finite differences, parameter shift or even stochastic versions of the parameter shift technique [13].

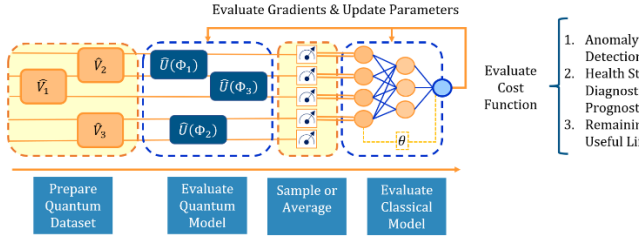

Figure 2: Hybrid Quantum-Classical Machine Learning approach design to perform a variety of tasks within the prognosis and health management field.

## 4 CASE STUDY: MFPT DATASET

### 4.1 MFPT Dataset and Processing Steps

The proposed framework is tested on the well-known ball bearing dataset generated by the Machinery Failure Prevention Technology Society (MFPT). For this dataset, a ball bearing test rig is subjected to three different health conditions with varying levels of load applied, while accelerometer data is recorded. First, 3 ball bearings with no faults and applied loads of 270 lbs. are tested as a baseline. Secondly, 10 ball bearings with faults located in the outer ring are subjected to 10 different levels of the applied load. Finally, 7 ball bearings with a fault located in the inner ring are tested for 7 different load levels. Table 1 presents a brief description of the ball bearings dataset.

Table 1: Brief description of the MFPT ball bearing dataset.

|  | Baseline (No Fault) | Outer ring (constant load) | Outer ring (varying load) | Inner ring (varying load) |
|---|---|---|---|---|
| Number of ball bearings | 3 | 3 | 3 | 7 |
| Load levels [lbs.] | 270, 270, 270 | 270, 270, 270, | 25, 50, 100, 150, 200, 250, 300 | 0, 50, 100, 150, 200, 250, 300 |
| Sample rate | 97656 [Hz] | 97656 [Hz] | 48828 [Hz] | 48828 [Hz] |
| Shaft rate | 25 [Hz] | 25 [Hz] | 25 [Hz] | 25 [Hz] |
| Measurements | 6 [s] | 6 [s] | 3 [s] | 3 [s] |

The original accelerometer signals were first down sampled to a uniform sampling rate of 48828 [Hz] and then split into segments of length $L = 4000$ points, with an overlap between adjacent segments of 200 points (5%). While the outer race fault ball bearings can be separated into two sub-classes (constant or varying load), for this case study we join them and consider three classes: baseline (B), inner ring fault (IR) and outer ring fault (OR)

For each segment of 4000 points, we extract 5 features: average, variance, maximum amplitude, peak-to-peak and root mean square. After this step, we obtain a dataset of 1,456 samples, which we divide into 80% and 20% for training and testing, respectively.

### 4.2 Quantum Data Encoding

Once the classical dataset is generated from the raw accelerometer data, it is required to encode it into objects that the quantum components of the algorithm can read and operate with. These objects are circuits in a bijective relationship with the datapoints. That is, one datapoint has exactly one circuit representation. The encoding is performed used *Angle Encoding* [10], which was discussed in Section 2.2.

### 4.3 Model Specification

For this case study, it is required to specify both the PQC and traditional neural network that compose the hybrid model. While the PQC can be chosen to have an arbitrarily complex inner structure (similar to what happens with the architecture selection process in deep learning), for this paper we will rely on performing rotations in all axes, without superposition. This will allow the system to freely shift the qubits orientation and phase, while maintaining the computation cost low. The PQC diagram can be found in Figure 3 (left), where it can be seen that every input qubit is passed through rotation gates for the Y, X and Z axes. The parameters $[\alpha, \beta, \gamma]$ are optimized during the training phase. Finally, a z-measure operation is applied to each qubit to convert the quantum data back to a classical configuration. The traditional neural network, Figure 3 (right), consumes as input this classical information and processes it with a three-layer configuration of 5 input units, 10 hidden units and finally, 3 output units that in conjunction with a softmax layer gives a score for each class, that indicates the model diagnosis of the datapoint.

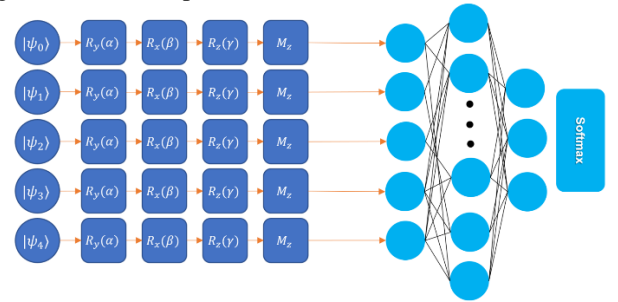

Figure 3: Hybrid Quantum-Classical Model Architecture; Left model shows the quantum gates selection for the PQC, with free parameters $[\alpha, \beta, \gamma]$ and z-measure gates for each qubit; right model shows the architecture selection for the diagnosis network, which is a feed-forward neural network with an input layer of 5 inputs, a hidden layer of 10 units and an output layer of 3 units in conjunction with a softmax activation function. The activation function for the inner layers is the elu function.

The model is trained using a categorical cross-entropy loss function. The optimization itself is implemented using backpropagation, considering the PQC as a black box and therefore using finite differences [13] to compute an approximation of the gradients that correspond to the parameters $[\alpha, \beta, \gamma]$.

The model is trained and evaluated 25 times, each with a different seed to present average and deviation results. The

Adam [16] optimizer is used, with a learning rate of 0.01 and 150 epochs per training. It is executing using Python 3.8.10, TensorFlow 2.4.1 and Tensorflow-Quantum 0.5.0 on a machine with an Intel i5-7300HQ processor and 16 GB of RAM. No GPU is used for the training of this model. The code and instructions required for reproducing the results of this paper will be released in GitHub, using docker containerization to assure correct replicability.

### 4.4 Results

As mentioned before, results are obtained by training the model 25 times to obtain averages of the metrics. Figure 4 and Figure 5 shows representative curves for the accuracy and loss evolution during training, respectively. As it can be seen, the model rapidly reaches a state of local minima, from which further improvements are harder to accomplish. Overall, the curves show a clear indication that the hybrid quantum-classical model obtains good results that leverage on the features extracted in the quantum domain, i.e., the quantum stage works as a feature identification and construction from the input signal, and this is performed in a higher-dimensional complex Hilbert space. Table 2 shows that the model does not overfit significantly during the training phase, as portrayed by the similar scores for the loss and accuracy metric obtained with the training and testing dataset once the model has finished its learning phase. Standard deviation measures show that the model trains consistently over random parameter initialization and stochastic backpropagation through both the classical neural network and the parameterized quantum circuit.

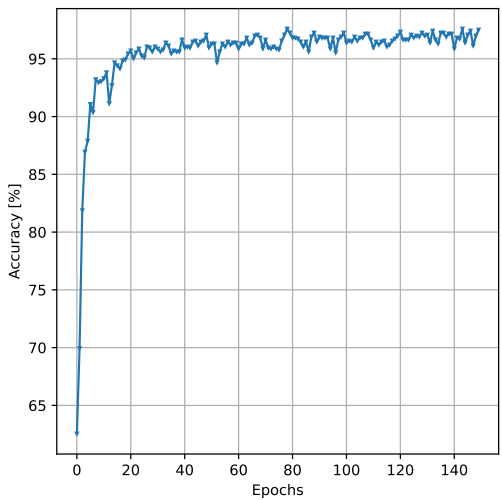

Figure 4: Accuracy evolution during the training phase.

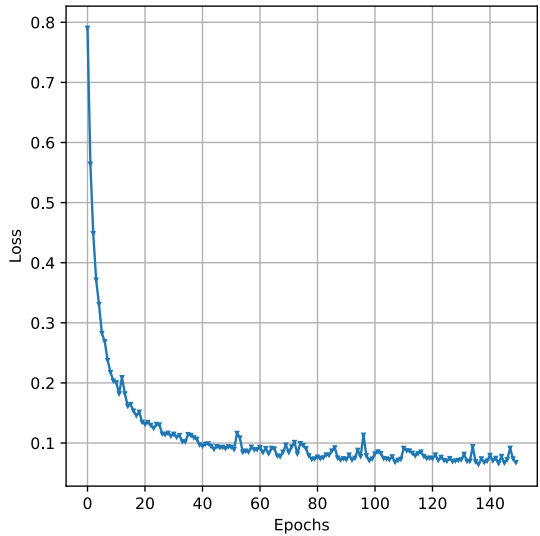

Figure 5: Categorical cross-entropy loss during the training phase.

Table 2: Performance metrics for the hybrid model. All results are obtained running the model 25 times and averaging the individual run's results. Format: (mean) ± (std).

| Metric | Value |
| --- | --- |
| Training accuracy | $97.6 \pm 1.2\%$ |
| Testing accuracy | $97.4 \pm 1.3\%$ |
| Training loss | $0.06 \pm 0.03$ |
| Testing loss | $0.073 \pm 0.03$ |

Figure 6 shows the confusion matrix for the testing dataset. From this graphical representation of the classification accuracy per class, it is clear that the model generalizes well for all classes, even though there are health states that are easier to classify than others. For example, the inner race fault is very distinct in its representation than the outer ring faults, which occasionally gets mistaken by the no damage state. This could be caused by the relatively simple classical preprocessing that was performed on the raw accelerometer data, which consists of only 5 temporal features. Nevertheless, from these results one can argue that the hybrid quantum-classical model delivers competitive performance in identifying health states in ball bearings.

*Figure 6: Confusion matrix results for the testing dataset. ND: no damage class, OR: outer ring damage, IR: inner ring damage.*

## 5 CONCLUDING REMARKS

This paper presented a hybrid quantum-classical machine learning framework for fault diagnosis and prognosis that harnesses the quantum parameterized circuit ability to identify and construct features in a high dimensional complex Hilbert space, which then are passed to a classical neural network to perform health state classification. It was also shown that these types of models' construction and training can be simulated on normal, classical hardware and therefore the PHM community can start experimenting and founding new and useful ways of integrating quantum computing techniques into traditional diagnosis and prognosis schemas. The results also show that the inclusion of a PQC and the quantum data encoding schema is not detrimental towards performance or accuracy, and therefore it is a first step towards searching for ways to improve current PHM algorithms with quantum computing. The framework presented here and initially developed by TensorFlow Quantum is general and therefore, can be used not only for diagnosis tasks, but also for prognosis challenges such as remaining useful life prediction. The main objective of this paper is to encourage the PHM community to start exploring this field, as we believe it holds many potential advances in the speedup of online PHM, the enhancement of reliability and safety analysis of complex systems by taking advantage of the simultaneous evaluation of multiple scenarios possible with the entanglement of multiple qubits, and intrinsic uncertainty quantification and propagation due to the data processing in the quantum domain. These are current research topics pursued by the authors and open questions and opportunities for the reliability, safety and risk community.